\documentclass[a4paper,11pt]{article}
\pdfoutput=1 

\usepackage{jcappub} 
\usepackage{ulem}

\usepackage[T1]{fontenc} 
\usepackage[utf8]{inputenc}

\title{\boldmath Dark matter from primordial black holes would hold charge}

\author[a,b]{I. J. Araya,}
\author[c]{N. D. Padilla,}
\author[d,e]{M. E. Rubio,}
\author[f]{J. Sureda,}
\author[g]{J. Maga\~na}
\author[h]{and L. Osorio}

\affiliation[a]{Instituto de Ciencias Exactas y Naturales (ICEN), Universidad Arturo Prat, Avenida Arturo Prat Chac\'on 2120, 1110939, Iquique, Chile}
\affiliation[b]{Facultad de Ciencias, Universidad Arturo Prat, Avenida Arturo Prat Chac\'on 2120, 1110939, Iquique, Chile}
\affiliation[c]{Instituto de Astronomía Teórica y Experimental (IATE), CONICET-UNC, Laprida 854, X5000BGR, Córdoba, Argentina}
\affiliation[d]{Scuola Internazionale di Studi Avanzati (SISSA), Via Bonomea 265, 34136 Trieste, Italy}
\affiliation[e]{Institute for Fundamental Physics of the Universe (IFPU), Via Beirut 2, 34014 Trieste, Italy}
\affiliation[f]{Donostia International Physics Center (DIPC), Paseo Manuel de Lardizabal, 4, 20018 Donostia-San Sebastián, Spain}
\affiliation[g]{Escuela de Ingenier\'ia, Universidad Central de Chile, Avenida Francisco de Aguirre 0405, 171-0164 La Serena, Coquimbo, Chile}
\affiliation[h]{Facultad de F\'isica, Pontificia Universidad Católica de Chile, Vicuña Mackenna 4860, Santiago, Chile}

\emailAdd{ignaraya@unap.cl}

\abstract{
We explore the possibility that primordial black holes (PBHs) contain electric charge down to the present day. We find that PBHs should hold a non-zero net charge at their formation, due to either Poisson fluctuations at horizon crossing or high-energy particle collisions. Although initial charge configurations are subject to fast discharge processes through particle accretion or quantum particle emission, we show that maximally rotating PBHs could produce magnetic fields able to shield them from discharge. Moreover, given that electrons are the lightest and fastest charge carriers, we show that the plasma within virialised dark matter haloes can endow PBHs with net negative charge. We report charge-to-mass ratios between $10^{-31}\,C/\mbox{kg}$ and $10^{-15}\,C/\mbox{kg}$ for PBHs within the mass windows that allow them to constitute the entirety of the dark matter in the Universe.} 

\begin{document}
\maketitle
\flushbottom

\section{Introduction}

The nature of Dark Matter (DM) in the Universe still remains unknown, even after over four decades since its discovery \citep{Rubin1978}. There exists a large amount of confirming evidence gathered since then, both in individual example studies such as the Bullet Cluster \cite{bullet}, in statistics of clustering of galaxies \cite{eboss}, in the abundance of galaxy clusters \cite{clusterabundance}, and imprints in the fluctuation spectrum of the Cosmic Microwave Background (CMB, \cite{Planck:2018vyg}).

Possible DM candidates include a wide variety of particles from extensions of the Standard Model of Particle Physics \cite{Feng:2010}, which are being looked for with several ongoing experiments \cite{Bernabei08,Mack07,Albuquerque10,Drukier86}.  On the modelling side, the expected effects of DM on different astrophysical observables are being intensely studied using numerical simulations (Fuzzy DM \cite{Hu}, Warm DM \cite{Viel:2005} and cold dark matter (CDM) \cite{Davis:1985}).  Among these, there are a few candidates that could hold electrical charge, such as the mini-charged DM particles \cite{Munoz:2018}. The effects of having electrically charged DM have been studied in the past with particular interest, and there are strong constraints on the charge per mass that could characterise them \cite{Caputo:2019}.

On the other hand, DM could also be constituted by compact objects, and in particular by black holes (BHs). Given that DM is present at least since the Early Universe \cite{Planck16,Planck:2018vyg,Arbey:2021gdg,Khlopov_2006}, it should necessarily have a primordial origin. This motivates one fundamental question driven in this paper, on whether or not primordial black holes (PBHs), as legitimate DM particle candidates \cite{Carr2020,Hawking1971,Bai2020,Khlopov_2010}, could also contain electric charge, and how one could use constraints on the charge for DM particles to turn these into constraints for charged PBHs as dark matter.

From an astrophysical view, the understanding of black hole physics and the role of black holes in a wide spectrum of physical phenomena with different scales has considerably grown with Numerical Relativity and cosmological simulations. For instance, the development of new techniques for evolving black hole binary solutions, as well as their interaction with electromagnetic fields, allowed for a more accurate modelling of new candidates for the generation of the recently detected gravitational waves \cite{Sperhake:2008ga,Liebling:2016orx,Barack:2018yly}. As it is well understood, black holes have “no hairs”, being their dynamics completely determined by only three parameters: mass, spin and charge. While the first two parameters seem to play a more important role in the majority of astrophysical processes, the study of charged black holes in the context of Astrophysics has not been of much interest in the community, being often neglected and (implicitly) set to zero. Perhaps this assumption comes up due to the expectation that either: (i) black holes do not have electric charge or, (ii) if they do, it is very small. Some ways to argue the above hypothesis are, on the one hand, the relatively large strength of the electromagnetic force when compared to gravitational interactions, which implies that black holes formed from gravitational collapse should remain (nearly) neutral. On the other hand, it is often claimed that the presence of plasmas around black holes leads to a rather rapid discharging. Some previous estimates \cite{Gibbons75} even predict that the rate of electric discharge experienced by charged black holes turns out to be rather large if
\begin{equation}
\frac{\sqrt{ck_c}}{G}\frac{Q}{M} \gtrsim 10^{-13}\left(\frac{J}{M}\right)^{-1/2}\left(\frac{M}{M_{\odot}}\right)^{1/2},
\end{equation}
where $M$, $Q$ and $J$ are respectively the corresponding parameters of mass, charge and angular momentum per unit mass. Although “zero charge” may be a good approximation when dealing with photon dynamics or more generally with neutral particles, both classical and relativistic processes can lead to certain amount of charge for BHs. Theoretical calculations predict quite straightforwardly that, even a small charge can significantly modify the motion of charged particles, like cosmic rays, making thus the problem of black hole charge a highly non-trivial challenge. Thus, some natural questions appear along the way: Is the charge of “realistic” black holes always equal to zero? If not, how could black holes get charged? Could this property provide any observable/detectable effects? Some of these questions have been recently addressed in the context of charged BH binary numerical simulations \cite{Bozzola1,Bozzola2,Bozzola3}.

There exist several reasons why it would be relevant to study whether or not these electrically charged objects exist in nature, at which scales, and under which physical mechanisms they would be able to accumulate and even maintain an average charge. In the context of Particle Physics, it is known about the possibility of the formation of black holes as a result of highly energetic collisions, a process in which any other type of interaction (such as electromagnetic) should be completely irrelevant \cite{Choptuik10,BozzolaPRL22}. However, recent studies on shock wave collisions suggest that the physical properties of the black holes resulting from such collisions, as well as their electromagnetic emission, do depend on the charge of the particles before colliding \cite{Lehner12}. Of course, this conjecture could be directly verified by comparing numerical simulations of charged particle collisions with those coming from electrically neutral systems. From the observational point of view, the presence of charge within astrophysical or even primordial BHs could be measured via multi-wavelength electromagnetic-based observations. In particular, experimental data from observations of Sagittarius A* cannot rule out the presence of electric charge within it \cite{EHT22a,EHT19,EHT16}.

In the context of astrophysics, there are theoretical estimates that favour the existence of charged black holes. A seminal work by Wald suggests that a rotating black hole in an external magnetic field (MF) induces an electric field due to the twisting of magnetic field lines \cite{Wald74}. 
Following Wald’s argument, the value of such an induced charge is proportional to both the strength of the magnetic field and the spin of the black hole. The sign of such a charge will depend on the orientation of magnetic field lines with respect to the black hole spin.  Another interesting proposal for generating electromagnetic counterparts coming from binary black hole mergers recently appeared, and has one intriguing requirement:  at least one of the binary components must have a non-zero amount of electric charge \cite{Zhang16}.  It has been argued that a charged black hole with non-zero angular momentum per unit mass may carry a “force-free” magnetosphere which allows the maintenance of electric charge for a quite long period of time \cite{Carrasco21,Carrasco18,ReulaRubioFF}. Finally, several studies on evolution of charged black holes in Numerical relativity have been relevant during the last decade \cite{Zilhao14,Lehner12,Liebling:2016orx,Hirschmann18,Kritos22}.

Another natural question that arises in this context is whether or not  primordial black holes could also accrete charge during their formation and subsequent evolution, and if so, if they follow a similar mechanism to the one followed by stellar or supermassive black holes. In particular, modelling charge accretion mechanisms in PBHs might give clues about the electrical state of those PBHs that have not yet evaporated, as well as point to whether dark matter formed by them may or may not be endowed with electric charge.

In this work, we study different mechanisms that would charge PBHs, or prevent their discharge. We assume initial charge configurations from Poisson charge imbalance within overdense super-horizon regions collapsing at horizon-crossing, or due to head-on collisions of high-energy charged particles during reheating. We see how the initial charge is completely lost by the time of decoupling, unless there are further processes that slow-down the discharge. We show that PBH could get charged by diffusion processes within late-time dark-matter haloes; since electrons are the lightest and thus fastest charge-carrier, their final net charge should be negative. For massive PBHs, we account for the build-up of baryons and pressure gradient in the vicinity of the BHs.  We finish by exploring the role played by the magnetic fields generated from rotating PBHs in slowing-down the PBH discharge.

\subsection{Outline and conventions}

This paper is organised as follows. In Section \ref{sec:accretion} we set up possible initial charge configurations as PBHs enter the horizon or form via particle collisions. In Section \ref{Sec:3} it is shown how low-redshift PBHs within DM haloes could acquire non-zero negative charge, even if the initial charge is completely lost. In Section \ref{sec:discharge}, we discuss three different mechanisms that could rapidly discharge PBHs. In Section \ref{sec:shielding} we show that for maximally rotating PBHs, sufficiently strong magnetic fields could be generated, preventing the loss of their initial charge. A brief discussion of our main results is given in Section \ref{sec:discussion}. Finally, the conclusions are presented in Section \ref{sec:conclusions}.

Throughout this work, we assume a flat cosmology with $\Omega_{m,0} = 0.315$; $\Omega_{dm,0} = 0.264$; $\Omega_{r,0} = 9.237 \times 10^{-5}$ and $H_0 = 67\, \frac{\mathrm{km}}{\mathrm{s}} \mathrm{Mpc}^{-1}$, in consistency with the latest measurements from the Planck Satellite \citep{Planck:2018vyg}. Also, we use the international system of units and, as usual, $e$ is the absolute value of the electron's charge, $c$ is the speed of light in vacuum, $k_c$ is Coulomb's constant, $G$ is Newton's universal constant, $h$ is Planck's constant, $k_B$ is Boltzmann's constant and $\varepsilon_0$ and $\mu_0$ are the vacuum electric permittivity and magnetic permeability, respectively.

\section{PBH charge at formation}
\label{sec:accretion}

In this section we discuss two possible ways for PBHs to form with non-zero initial charge; namely, (i) from Poisson fluctuations in the number density of charged particles within the volume that enters the cosmological horizon, and (ii) from high energy collisions of charged particles in the Early Universe. We then show that accretion of charged particles could discharge PBHs very early on. These processes are insensitive to the fraction of DM in the form of PBHs.

\subsection{Initial charge from Poisson fluctuations}

Even though any initial PBH charge is expected to be lost quite quickly we will argue in this paper that under certain conditions, this loss can be prevented for maximally spinning PHBs.  Therefore, here we estimate the initial conditions for  the PBH charge. We follow a similar argument to that in \cite{Maldacena,araya20} for a non-zero magnetic monopole charge in PBHs, due to the stochasticity of the abundance of particles in patches that have not yet come into causal contact, i.e. Poisson noise in the number of particles that become entrapped in the black hole as it forms.  This initial charge, which can be of either sign, is specific to the  horizon crossing formation scenario \cite{Sureda:2020}.  

We assume that at the moment when the PBH enters the horizon, it contains a Poisson excess of charge of either sign. Since the collapse threshold of primordial black holes is of order $1$ (see \cite{Sureda:2020} and references therein),  we simply calculate the temperature $T$ at which the volume of the black hole equals that of the Horizon, i.e. $c/H(T)=V^{\frac{1}{3}}$, where we use the approximate value of the expansion rate of the universe prior to matter-radiation equality, $H(T)\sim H_0 \sqrt{\Omega_{r,0}} (T/T_0)^2$, $H_0$ and $\Omega_{r,0}$ are the present-day Hubble parameter and radiation energy density, respectively, $V$ is the volume of the black hole,  $V=M/\rho_{\rm tot}$, where $M$ is the black hole mass and $\rho_{\rm tot}\simeq \rho_c \Omega_{r,0} (T/T_0)^4$ is the total energy density at temperature $T$, $T_0=2.73K$ is the temperature of the cosmic microwave background, and $\rho_c$ is the critical density of the Universe. We adopt an upper limit of $T<10^{17}$GeV, which is our assumed reheating temperature, since the number density of any PBHs  forming prior to the end of Inflation would be washed out exponentially.  This sets the lower limit of PBH mass formed at horizon entry to $M\simeq 10^{-4}$kg (this limit is shown in Fig. \ref{fig:RN+Poiss} as a vertical line).  The resulting Poisson charge can be approximated by,
\begin{equation}\label{eq:Q0poisson}
Q_{\rm init}^{\rm Poisson}(M) = \pm e \sqrt{\max{(1,V n_e(T(M)))}}.
\end{equation}
The number density of electrons, $n_e$, is taken to be that of the baryons up to the electron-positron annihilation at $T_{\rm ann}\sim 16\,\mbox{keV}$.  At higher temperatures we simply multiply this by the inverse baryon to photon ratio $\eta$ to account for the increased number density of charges when these particles were in equilibrium with the radiation.  This is of course only approximate as it ignores the other charged particles at very high $T$, but it is useful for the order of magnitude calculations we will perform here. We take the maximum between 1 and the number of charges within $V$.

The resulting initial Poisson charge as a function of the PBH mass can be appreciated in Fig. \ref{fig:RN+Poiss}, which as can be seen increases as $M^{3/4}$ up until roughly $M=10^{38}\,\mbox{kg}$, where the electron-positron annihilation decreases drastically the number of available charges when such large PBHs form as they enter the horizon, after this particular decoupling.

Notice that, in the whole temperature region of interest (and down to PBHs of Planck mass), {$Q_{\rm init}^{\rm Poisson}$} is always several orders of magnitude lower than the charge that satisfies the limit {imposed by the condition for the PBH not to be superextreme  \cite{Wald:106274}, namely,
\begin{equation}\label{extremal-limit}
    2r_Q \leq r_{\text{s}},  
\end{equation}
where
\begin{equation}
    r_{\text{s}}=\frac{2GM}{c^{2}}, \quad 
    r_{Q}=\sqrt{\frac {k_cGQ^{2}}{c^{4}}},
\end{equation}
shown by the red line in the same figure.}

\subsection{Planck mass PBHs with extremal charge}
\label{sec:rnextremal}

In this subsection we discuss another possible mechanism for forming charged PBHs, which could come from high-energy collisions of electrons or positrons in the very-early universe, and around the time of reheating. As presented in \cite{Carneiro:2020}, the head-on scattering of electrons could form charged PBHs with \textit{extremal} electric charge, provided that the center-of-mass energy of the electrons is close to the Planck energy scale (we also refer the reader to the discussion pointed out in \cite{Pigozzo:2021}). The possibility of such high kinetic energies for electrons at the reheating scale seems implausible, as its energy is \textit{less} than the Planck scale. However, as shown in \cite{Carneiro:2021}, for reheating models occurring at high enough temperatures, $\sim 10^{17}$GeV, the high-energy tails of the particle velocity distributions could lead to enough collisions to produce PBHs so as to explain a significant portion of the DM. It is important to mention that said PBHs should be \textit{extremal} (i.e., maximally charged PBHs), in order for them to be stable under decay by Hawking radiation \cite{Hawking75}.

One possible way to understand how the charge of the electron could form an extremal PBH is as follows\footnote{We are grateful to Saulo Carneiro for providing us this argument.}. The maximum charge that an extremal PBH formed by the head-on collision of two electrons can have is the one which would make the electrons lose \textit{all} of their kinetic energy at the expense of the Coulomb potential energy, precisely at the gravitational radius associated to the mass of the BH. If the electric charge was more than this quantity, then the electrons would bounce off of each other, without forming the BH. Then, if the total energy of the system was equal to the kinetic energy of the electrons at infinity, this energy divided by $c^2$ would equal the mass of the BH. The total energy of the system would also equal the Coulomb energy at the gravitational radius, which allows to solve for the charge of the extremal PBH as a function of its mass. Then, considering the mass of the PBH as equal to the Planck mass, it gives a maximum charge of
\begin{equation}\label{qmax-saulo}
    q_{\text{\tiny{max}}}\sim\sqrt{\frac{G}{k_c}}m_{p}.
\end{equation}
This should only be considered as an order-of-magnitude argument, as it approximates the potential energy stored by the two electrons as given only by the Coulomb energy. 

A more rigorous way to find the charge required to make an extremal Reissner-Nordstr\"om (RN) PBH would be to simply look at the corresponding condition from the metric. This gives, precisely, a charge value given by Eq. (\ref{qmax-saulo}). However, even this calculation is approximate as \cite{Pigozzo:2021} shows, using an effective model, that there is a small but non-zero correction to the horizon of a Planck scale RN black hole. 

\begin{figure}
    \centering
    \includegraphics[width=1\textwidth]{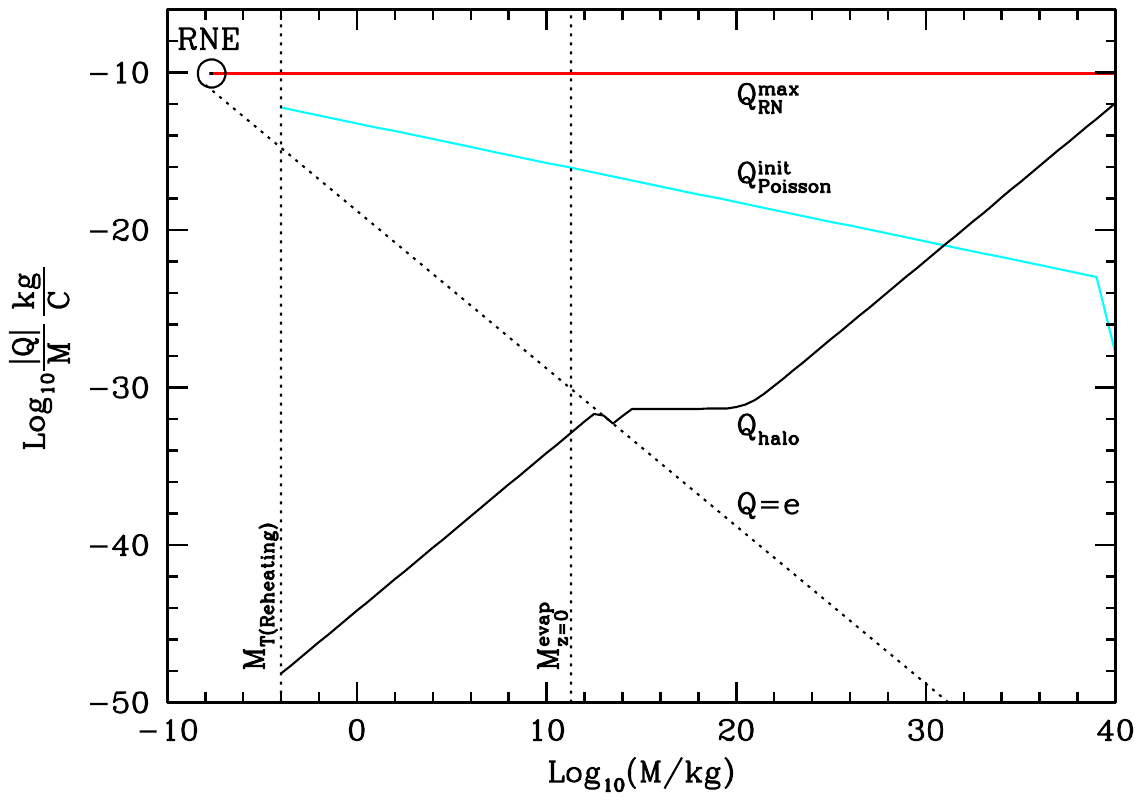}
    \vskip -7cm
    \caption{
    Absolute value of charge per mass for the initial Poisson charge (cyan line) and the extremal limit (\ref{extremal-limit}, red  line) as a function of PBH mass, for reference.  The RN extremal PBH is shown by the black open circle. Black dotted lines show the case of one electron charge, the mass of PBH formed right after reheating by collapse at horizon entry, and the mass of PBHs evaporated by the present-day (as indicated). The black solid is for the negative charge of PBHs within dark matter haloes at $z=0$.  }
    \label{fig:RN+Poiss}
\end{figure}

In any case, it is interesting to note that the obtained charge is close to the charge of the electron, as derived from considering the definition of the Planck mass and the fine-structure constant $\alpha$ \cite{Carneiro:2020}. Although both charges are not the same, as they differ precisely by $\sqrt{\alpha}$, which is a factor of around $1/10$. However, because there are two electrons involved in the collision which forms the BH, the \textit{miss-match} is reduced to a factor of 5. Indeed, said mismatch could be reduced even further when the running of the electric charge with the energy scale of the collision is taken into account. In particular, it is well-known that the effective electric charge of a particle involved in a scattering process changes with the center-of-mass energy of the collision, as it is seen in standard accelerator experiments \cite{Peskin95}. Then, the relevant center-of-mass energy scale for the PBH formation process would precisely be the Planck scale. If the electric charge was boosted by a factor of five at the Planck scale, due to the running of the electromagnetic coupling, then this mismatch factor could be explained. It is true that the latter point is purely ``speculative'', as the running of the electromagnetic coupling is not known beyond current accelerator scales. But the trend is for its magnitude to increase with energy, and so it is a reasonable extrapolated trend.  We illustrate these PBHs as an open circle in Fig. \ref{fig:RN+Poiss}. Note that this argument is also valid for positrons, making the average charge of the population of PBHs formed this way null.

Since the Hawking temperature for extremal RN PBHs is null, then these PBHs would not be subject to discharge by Hawking emission.  However, it is very likely that these extremal black holes are subject to accretion of charges from their surrounding plasma. They would also be subject to the Schwinger effect as discussed for instance in \cite{Chen2012} and references therein, but to date it is not exactly known how this discharge would occur. 


\section{Charge accretion from the surrounding plasma}\label{Sec:3}

The aforementioned charging mechanisms would result in a population of initially-charged BHs. However, due to the particle emission and accretion processes (cf. Section \ref{sec:discharge}), it is expected that this initial charge will be rapidly lost, except for cases where either extremality or some unknown shielding mechanism could prevent or slow-down the discharge (cf. Section \ref{sec:shielding}). Even though the initial charge may be lost, there is a mechanism able to charge PBHs in late-time virialized DM haloes, which will be discussed in what follows. 

\subsection{Charge accretion in late-time {DM} haloes}\label{Sec:halos}

After recombination of electrons and protons at $z\sim1100$, there are essentially no free charges to continue the evolution of the PBH charge by accretion.  This situation changes at reionization by the UV emission of the first stars, but the resulting plasma is of much lower density and temperature than at early epochs.  However, the collapsed structures dominated by dark matter in the late Universe heat baryons up to the virial temperature, which can reach temperatures in excess of $10^7$K, for instance, in galaxy clusters \cite{Voit}.

We propose a simple model for the rate of charge accretion of BHs within the environment of virialised dark matter haloes by assuming that electrons and Hydrogen nuclei are in thermal equilibrium at the virial temperature of haloes.  We remind the reader that PBHs constitute extremely cold dark matter that allow the formation of virialised dark matter haloes of masses of a few orders of magnitude higher than that of the highest mass PBH.
A first approximation for the resulting rates of charge accretion can be obtained by adopting an overdensity $\delta_c$ of virialised structures of $\delta_c=200 \rho_c$, where $\rho_c$ is the critical density of the Universe, and neglecting the presence of accretion discs in PBHs.

It is possible to take a heuristic approach to estimate the average charge of the population of PBHs in haloes.  We will concentrate first in the case when at most, a PBH will acquire a charge of $Q=e$ within a Hubble time via diffusion, from its surrounding plasma; this choice will be justified shortly.   
Assuming the cross-section for neutral PBH charge accretion as the geometrical one at the innermost stable circular orbit, $\sigma_{\rm PBH}=\pi (1.4 r_+)^2$, 
with $r_+$  the radius of the exterior horizon of the RN metric, namely
\begin{equation}
    r_+=\frac{r_{\text{s}}+\sqrt{r_{\text{s}}^2-4r_{Q}^2}}{2},
\end{equation}
the rate of accretion by PBHs of mass $M$ reads \cite{longair1998galaxy},
\begin{equation}
    \Gamma=n_e^h \sigma_{\rm PBH}  \left(v_e(T_{\rm virial})-v_p(T_{\rm virial})\right),\label{eq:gammaH}
\end{equation}
where the number density of electrons in the halo is approximately $200$ times the cosmic average, $n_e^h\sim 200 n_e$.
Electrons and atomic nuclei are assumed to be in thermal equilibrium at the virial temperature $T_{\rm virial}=GM_hm_p/(3k_BR_h)$ (or $T_0$, whichever is the highest) within haloes of mass $M_h$ and radius $R_h=(3M_h/(4\pi 200 \rho_c))^{1/3}$, as well as the black holes conforming a fraction $f\leq1$, of the dark matter that makes up most of the mass of such haloes.  This rate is lower than the present day Hubble parameter, $H_0$, for $M<10^{13}$kg, justifying ignoring details such as the Debye length of the plasma within this PBH mass range.  The average charge after a Hubble time will simply be the fraction of PBHs with $Q=e$ times $e$, namely
\begin{equation}
    \bar{Q}=-\frac{\Gamma e }{H_0}\label{eq:fraccharge},
\end{equation}
and is shown as a black line in the left segment of Fig. \ref{fig:RN+Poiss} for $M_h=10^{15}M_\odot$.
The resulting charge is negative since the lower mass of electrons makes their thermal velocity larger than that of protons.

The vertical line at $M\sim10^{12}$kg marks the mass of Schwarzschild PBHs that should evaporate by $z=0$ due to Hawking radiation.  We bear this limit in mind but still estimate the charge and related effects for lower mass black holes for completeness.

For higher mass PBHs with $M\gtrsim10^{13}$kg, where the rate of accretion is larger than $e$ within a Hubble time, we make two heuristic assumptions.  The first being that the acquired charge cannot exceed the total charge that is able to enter a Debye volume, $V_D$ within a Hubble time,
\begin{equation}
    Q_{D}=\frac{e n_e V_Dv_p}{\ell_D H_0},
\end{equation}
where $V_D=\sigma_{\rm PBH} \ell_D$, and the Debye length is,
\begin{equation}\label{eq:Debye-lenght}
    \ell_D=\frac{k_B T_{\rm virial}}{4 \pi k_c e \sqrt{n_e}}.
\end{equation}
Note that as the Debye volume involves the cross-section of the PBH, the Debye length cancels out. However, this choice stresses the plasma nature of the intra halo medium, where the bulk velocity of charges is well approximated by that of the slowest charged particle.
As can be seen in Fig. \ref{fig:RN+Poiss} this condition limits the charge to $Q=e$ around $M=10^{13}$kg, but the cross-section of larger PBHs allows $Q_D$ to increase and, in turn, the charge increases again just at slightly larger masses.  Here we are ignoring the Coulomb repulsion force exerted by the PBH on the electrons (the fast particle which contributes charge to the PBH) and the accretion of the protons (which are 30 times slower than the electrons within a Debye length of the BH and can therefore be ignored), which implies that this is likely an overestimation of the charge.

The second heuristic assumption is that electron accretion can proceed only until the negative PBH charge repels electrons to the point where they lose all their kinetic energy when approaching the PBH from a distance of one Debye length to the outer horizon of the PBH, taking into account the black hole charge, $Q$. This gives the final charge for a PBH, as no further electrons can be accreted due to Coulomb repulsion. This translates into,
\begin{equation}
    Q_{\rm max}=\frac{k_B T \left( \frac{1}{r_+}-\frac{1}{r_++\ell_D}\right)^{-1}+GMm_e}{k_c e}.
\end{equation}

This is shown as the leveling off of the black line in Fig. \ref{fig:RN+Poiss}, where the charge reaches an upper limit of $Q/M\sim 10^{-31}C/$kg, up until $M\sim10^{20}$kg. At this mass, the charge increases again as the Debye length becomes smaller than the outer horizon radius. Since the proton velocity is $\sim 2$ orders of magnitude lower than that of electrons within a Debye length, we ignored the in-fall of protons in this last equation. Therefore, our estimates of maximum charge are only slightly larger than what would result from a full calculation.  

The approximation presented here allows us to infer that the charge of PBHs in haloes would be of negative sign.  This implies that the Coulomb potential could be quite large as dark matter would not be neutral on average.  The condition for the potential energy to be dominated by gravity is the same as for the extremal charge in the RN metric, i.e. $Q<\sqrt{G/k_c}M$. We limit our analysis to PBH masses where this condition is met.
These assumptions are reasonable as long as any charge evolution (either discharge from initial Poisson value, or charging from neutral) is a rather slow process.  

Up to this point we are ignoring the effect of the build up of baryons due to the gravity of the PBH; namely, we are using a diffusion process to account for the accretion of charges.  However, the PBH is able to capture baryons via gravity in an accretion region where there could be a stationary state in near hydrostatic equilibrium, such that the gravitational force is compensated by the pressure gradient, and thus, gravity can be neglected from Eq.(\ref{eq:Debye-lenght}). In this case the baryons arrange themselves in an overdensity around PBHs near the innermost stable circular orbit, where their density is enhanced. In order to assess qualitatively the changes in our estimates due to these effects, we simply adopt different density enhancement factors, and in particular, for a factor $10^{6}$ higher, the charge increases by $3$ orders of magnitude.  This indicates that our simple estimates for diffusion are lower limits for the actual charge in PBHs that conform dark matter haloes.

Having provided estimates for the initial charge of black holes in the early Universe, and for late times accretion of charge, we will now attempt to obtain some insight on how the charge could evolve between these two conditions.
To do so we take into account that PBHs are subject to accretion of charges from their surrounding plasma, and to production of electrons and positrons by Hawking radiation and via the Schwinger effect \cite{Gibbons75}. We study how these affect the final average charge in the next section.

\section{Discharge of black holes in the early and late Universe}\label{sec:discharge}

In this section we explore the processes that can lead PBHs to loose their 
charge, obtained either at formation for the RN extreme PBHs,  by enclosing unbalanced positive and negative charges due to Poisson noise as the volume of the PBH enters into causal contact and forms its event horizon, or at late times from the accretion of surrounding baryonic plasma in virialized structures.  
In the first subsection we look at the process of discharge by accretion, and in the second subsection, at discharge by pair emission processes.

\subsection{Discharge by particle accretion}

As in Section \ref{Sec:halos}, here we concentrate on the accretion of charges from the plasma surrounding PBHs considering only a diffusion process, ignoring the possible build up of baryon overdensities around PBHs; we have also taken into account the effect of restoring pressure in cases where an accretion region could build up around PBH and find that the following results remain virtually unchanged.  In order to do the calculation of discharge for
the case of PBHs with charge prior to decoupling of baryons and radiation, we compare the rate of particle accretion to that of the Hubble expansion, so as to infer the lowest temperature down to which PBHs could have lost their charge by particle accretion, $T_{\rm qdec}$. This temperature  will depend on the PBH mass and can be written as,
\begin{equation}
    n_{\rm e} \sigma_{\rm PBH} v_{\rm slow}(T_{\rm qdec})=H(T_{\rm qdec}),\label{eq:Tqdec}
\end{equation}
where the PBH cross section reads $\sigma_{\rm PBH}=\pi r_{E}^2$, and
$r_{E}$ is the radius at which the total energy is zero, 
\begin{equation}
    E=-\frac{ k_c|Q|e+GMm_{\rm slow}}{r_{E}}+3 k_BT=0;
\end{equation}  
unless it is lower than the innermost stable circular orbit in which case it is set to $r_E=1.4r_+$, or if this radius exceeds the separation at which the potential drops exponentially due to the plasma being neutral, in which case we set it to $r_E=1.4 r_+ +  \ell_D$.
This is valid for either positive or negative initial charge.
We assume that the electron number density is a reasonable approximation to the number density of charged particles at any temperature when taking into account their increased number density prior to the electron-positron annihilation as in Eq. \ref{eq:Q0poisson}. We use the thermal velocities of the slower particles $v_{\rm slow}$, which corresponds to the that of the proton at low temperatures, and to that of electrons/positrons above $T_{\rm ann}$  (we notice that above the QCD scale there are more massive charged particles, but this approximation is still valid as all particles are already relativistic).  

\begin{figure}
    \centering
    \includegraphics[width=0.9\textwidth]{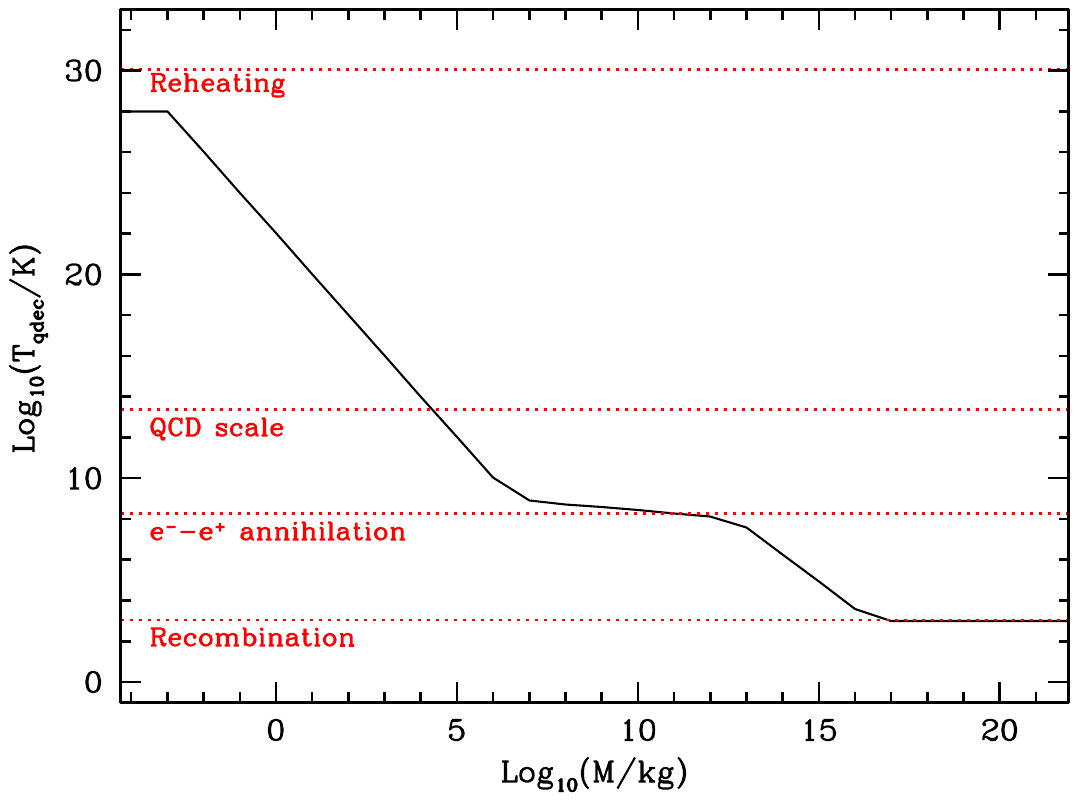}
    \vskip -7cm
    \caption{Temperature of charge decoupling as a function of PBH mass (black solid line).  The horizontal lines show the typical value for reheating adopted here of $10^{17}$GeV, the QCD scale, the temperature of electron-positron annihilation, and recombination.}
    \label{fig:decoupling}
\end{figure}

\begin{figure}
    \centering
    \includegraphics[width=1\textwidth]{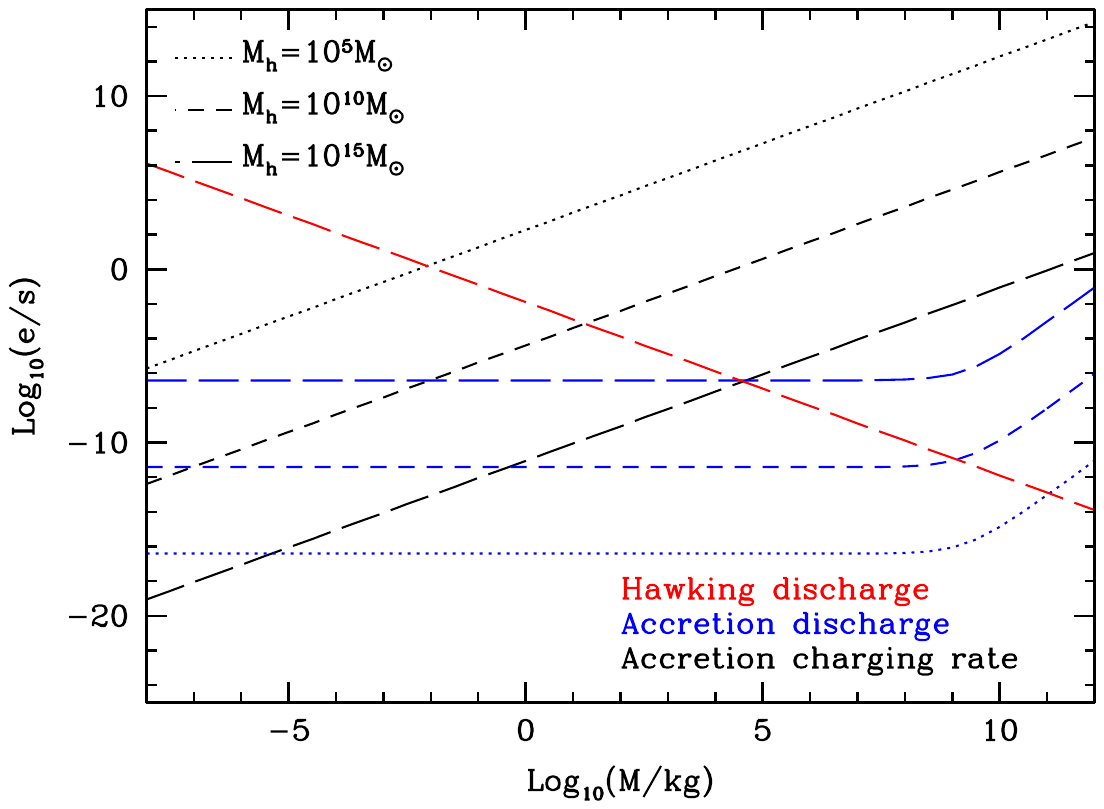}
    \vskip -7cm
    \caption{{ Rates of discharge by accretion of particles from the late-time plasma in dark matter haloes (blue lines), and via Hawking evaporation (red), for PBHs of charge $Q=e$ as a function of their mass, compared to the rate of charging of PBHs in haloes (black lines) of different masses as indicated by the legend.}} 
    \label{fig:rates}
\end{figure}

Our calculation for a decoupling temperature between charged particles and black holes would be justified if any charge evolution is slow, such that both positive and negative particles reach the PBHs at the pace of the slowest particles, which is expected as the fluid is a plasma where velocities are homogeneised by Coulomb scattering; this in turn justifies the use of the velocity of the slower particles, $v_{\rm slow}$, in Eq. \ref{eq:Tqdec}.  The resulting charge decoupling temperature as a function of PBH mass is shown in Figure \ref{fig:decoupling} for the initial Poisson charges (black solid line).  Regardless of PBH mass and initial charge sign, the decoupling with charges takes place at lower temperatures than that of horizon entry and PBH formation.

Therefore, PBHs should indeed discharge via accretion of particles { in the early Universe}.  Depending on the rate of discharge, it is possible that some charge could remain, particularly considering possible magnetic shielding against discharge by accretion (see Section \ref{sec:shielding}).

{ The case of loss of charge in dark matter haloes acquired at late times can also be approached this way, using the rate of Eq. \ref{eq:Tqdec} but comparing it to the rate with which PBHs in haloes of mass $M_h$ acquire a charge of $Q=e$ of Eq. \ref{eq:gammaH}, multiplied by the number of PBHs per halo $M_h/M$.  This comparison is shown in Fig. \ref{fig:rates} where it can be seen that the rates of discharge of PBHs with $Q=e$ are greater than the rates of charge (blue and black lines, respectively) for low mass PBHs. The different line types correspond to different halo masses (indicated in the figure), and this shows that for halo masses of $M_h=10^{15} M_\odot$ the rate of discharge is lower than the one for charging black holes of masses $M\sim10^5$kg.  This intersection takes place at lower black hole masses for more massive dark matter haloes.   }

\subsection{Discharge by particle emission}

It has been known for several decades that black holes can emit particles, such as charged electrons and positrons, via both Hawking radiation \cite{Hawking75} and the Schwinger process \cite{Gibbons75}.  In this section we study both effects to include them in our analysis.  The following is symmetric under a change of sign of the PBH charge.

\subsubsection{Hawking radiation}

Once the Hawking temperature is high enough compared to the mass of electrons and positrons, there is a chance that such particles will be emitted and cause the PBH to discharge.  There are several numerical solutions available for this problem, including public codes that can be used such as \url{BLACKHAWK} \cite{Arbey2019}, but for our purposes of order-of-magnitude calculations, we will resort to a simple estimate of the specific rate of discharge by electrons and positrons.

We start from the expression of ensemble average for a Bose-Einstein distribution function,
\begin{equation}
   \bar{X}=\int \frac{g(s)Vd^3p}{(2\pi \hbar)^3} x(p)f(\epsilon(p),\mu),
\end{equation}
where $\bar{X}$ is the ensemble average of a quantity $X$, $x(p)$ is the value of the corresponding quantity for a single particle with momentum $p$ and $g(s)$ is the spin degeneracy of the particle.

The distribution function reads,
\begin{equation}
    f(\epsilon(p),\mu)=\frac{1}{\exp{\left(-\frac{\epsilon(p)-\mu}{k_B T_{\rm H}}\right)}-1},
\end{equation}
where $T_{\rm H}$ is the Hawking temperature of the PBH,  $\epsilon(p)=\sqrt{p^2c^2+m_e^2c^4}$ is the kinetic energy of a single particle with momentum $p$ and mass $m_{e}$, and $\mu$ is the chemical potential that accounts for the Coulomb energy of the particle with electric charge $\pm e$ sourced by the PBH charge $Q$,
\begin{equation}
    \mu(\pm e)=\frac{k_cQ(\pm e)}{r_+},
\end{equation}
where $r_+$ is the outer horizon.

The charge density of created electrons or positrons in the vicinity of the BH horizon can be written as,
\begin{equation}
    \rho=Q/V=\int\frac{2Vd^3p}{(2\pi \hbar)^3}(\pm e) f(\epsilon(p),\mu)
\end{equation}
which allows to write the current as $\vec{J}=\rho\vec{v}$, from which we can write the time derivative of the charge as,
\begin{equation}
    \dot Q(\pm e)=\int \vec J(\pm e)\cdot d\vec A=A\int \frac{2d^3p}{(2\pi \hbar)^3} v(p) \cos(\theta(\vec p, \hat n)) (\pm e) f(\epsilon(p),\mu(\pm e)).
\end{equation}
Adding together the emission of electrons and positrons, regardless of the PBH charge sign, the final discharge rate due to Hawking emission reads
\begin{equation}
    -\dot Q= A\int_{0}^{\infty}\frac{dp p^2}{(2\pi)^2 \hbar^3} v(p) (- e)\left( f(\epsilon(p),\mu(-e))-f(\epsilon(p),\mu(e))\right).\label{Eq:Hawkdischarge}
\end{equation}

We will calculate the rate of discharge for $M\lesssim2\times10^{13}$kg which corresponds to the PBH mass that satisfies $k_BT_{\rm H}\sim m_e c^2$ and is susceptible to charge emission by Hawking radiation, for the Poisson initial charge and for the late time charge of PBHs in dark matter haloes.  

In the case of the initial charge, we adopt $Q=Q_{\rm Poisson}^{\rm init}$ and find that already at temperatures above the QCD scale, the Hawking rate is higher than the expansion one, implying that this mechanism could be able to discharge PBHs in this mass range from their initial Poisson charge.

In the case of late time charge of PBHs in haloes, this PBH mass range is almost entirely under the regime where  only a fraction of the PBH population contains a charge of a single electron.  Therefore, to determine whether the population of PBHs can be considered to hold a charge as an ensemble, we can use the accretion rate of individual PBHs of Eq. \ref{eq:fraccharge}, multiply it by the number of PBHs subject to acquire charge, and compare this to the rate of discharge of Eq. \ref{Eq:Hawkdischarge} assuming a PBH charge of $Q=e$.  The number of PBHs is obtained by simply assuming that dark matter in haloes is entirely composed of PBHs of equal mass $M$ { as in the previous subsection}. 

We show the comparison of the rates in haloes in Fig. \ref{fig:rates}.  
As can be seen, the Hawking rate of discharge drops steeply as a function of PBH mass, whereas the rate of charge by accretion increases, becoming larger than the rate of discharge for PBHs of mass $10^{-2}$ kg and $\sim 10^{5}$ kg for dark matter haloes of dwarf galaxies and typical clusters of galaxies respectively. Notice that $~10^{-4}$kg is the lowest mass that can form at horizon entry after our adopted temperature scale for reheating.  { Notice that for the halo masses shown here we can ignore the discharge by accretion of particles, as this rate of discharge  is always lower than either the Hawking rate or the rate of charging of PBHs, whichever of the two is the highest.   }

If dark matter were composed of PBHs, there would be a net negative charge per mass in dark matter haloes that would correspond to that of Section \ref{Sec:halos}, multiplied by the ratio between the Hawking discharge rate and the halo charging rate when the former is larger than the latter.  The resulting charge is shown, along with  the initial charge from Poisson processes in the Early Universe, and the maximal RN charge, in Fig. \ref{fig:final}, as  the black solid line for haloes of $M_h=10^{15}M_\odot$, and a dashed line for $M_h=10^{5}M_\odot$.

\subsubsection{Schwinger effect}

Another source of possible discharge of PBHs comes from the Schwinger effect \cite{Schwinger}, which proposes that a strong electric field can give rise to pair production of matter, in particular to charged particles that can quickly discharge black holes of up to $~10^{35}$kg ($\sim 5 M_\odot$).  The black hole charge that facilitates this process was worked out by \cite{Gibbons75} to be,
\begin{equation}
    Q^{\rm pairs}=\frac{2GMm_e}{k_c e},
\end{equation}
which results from the Klein Paradox evaluated at the Schwarzschild radius of a black hole.  This does not take into account the rate at which this production would take place.  However, \cite{Gibbons75} also provides an estimate of the rate by comparing the electric force to the one obtained by combining the rest mass and charge of the electron,
\begin{equation}
    Q^{\rm rapid}=\frac{4G^2m_e^2M^2}{k_c^2e^3}.
\end{equation}
When the charge of a BH approaches $Q^{\rm rapid}$ the discharge is efficient and the charge quickly drops to lower values.  

These two limits to the PBH charge are shown in Fig. \ref{fig:final} as dashed orange (pair production) and green (rapid) lines.  As can be seen, the Poisson initial charge is larger than the rapid discharge limit for $M\lesssim10^{24}$kg, and could therefore be lost early on due to this process.  On the other hand, the late time halo charge is below the rapid discharge value for all PBH masses which makes it stable under this type of charged pair production.  

Notice that the estimate of rapid discharge by \cite{Gibbons75} is only valid for large PBH masses, and cannot, in principle, be applied to low masses.  Therefore, we consider this process for PBHs with masses above $M\sim10^{12}$kg, which is also the range where PBH charges within dark matter haloes are $Q\geq e$, and where Hawking discharge is already unlikely.  

It is not clear whether this type of discharge process can affect the Planck mass extremal RN black holes of Section \ref{sec:rnextremal}, although as pointed above the discussion on this topic is still ongoing in the literature.

\section{Shielding accretion and emission of charges by PBHs}
\label{sec:shielding}

One possible way to discharge PBHs from an initial charge $Q_{\rm init}^{\rm Poisson}$, is from charge accretion of the opposite sign to the net charge that the PBH had initially. Here we will make an estimate of the effect on the trajectories of incoming charges onto charged, high spin PBHs.  The high spin of PBHs is justified since \cite{Mirbabayi20} show that PHBs could form with quite high spins. Here we do not attempt to solve this problem fully; instead, we give some simple estimates for this case and also extend it to the case of charged particle emission by Hawking radiation or via the Schwinger effect.

\subsection{Shielding against discharge by accretion}

A dipolar magnetic field surrounding a PBH could deviate charges incoming in radial trajectories toward the black hole.

It is well understood from first principles that electrically charged PBHs with nonzero angular momentum could produce a dipolar magnetic field.  Following \cite{Carter}, the magnetic moment reads $m=a Q c r_+$, and the magnetic field,
\begin{equation}
    B(r)=\frac{k_c a Q r_+}{c r^3},\label{eq:bspin}
\end{equation}
where $a$ is the dimensionless PBH spin.

This would constitute a dipolar field surrounding the black hole, which may deviate charges into non-radial trajectories. Considering the case in which the ratio between the electric and magnetic forces is an integer, say $F_E/F_B = N$, $N\in\mathbb{N}$, the radial distance $r$ to a relativistic charged particle following a radial trajectory satisfies
\begin{equation}
    \frac{r}{r_+}=Na.
\end{equation}
This would be effective at delaying the discharge of maximally spinning PBHs with $a\sim 1$, possibly allowing them to hold charges near the initial Poisson one of Section 2 at late times by avoiding accretion. We postulate another contribution to the shielding from magnetic monopole charges in Appendix \ref{app1}, which might add additional shielding for non-extremal spinning PBHs.

In a similar vein, black holes that are able to form accretion discs around them, i.e. those satisfying $M\gtrsim10^{12}$kg \cite{araya20}, would also produce magnetic fields via the Biermann battery mechanism that would effectively add to the shield against discharge by accretion.

\subsection{Shielding against discharge by emission}

If the charges emitted via Hawking radiation and the Schwinger effect can be prevented from escaping the PBH to infinity, this would make the PBH appear as electrically charged from afar.

\begin{figure}
    \centering
    \includegraphics[width=1\textwidth]{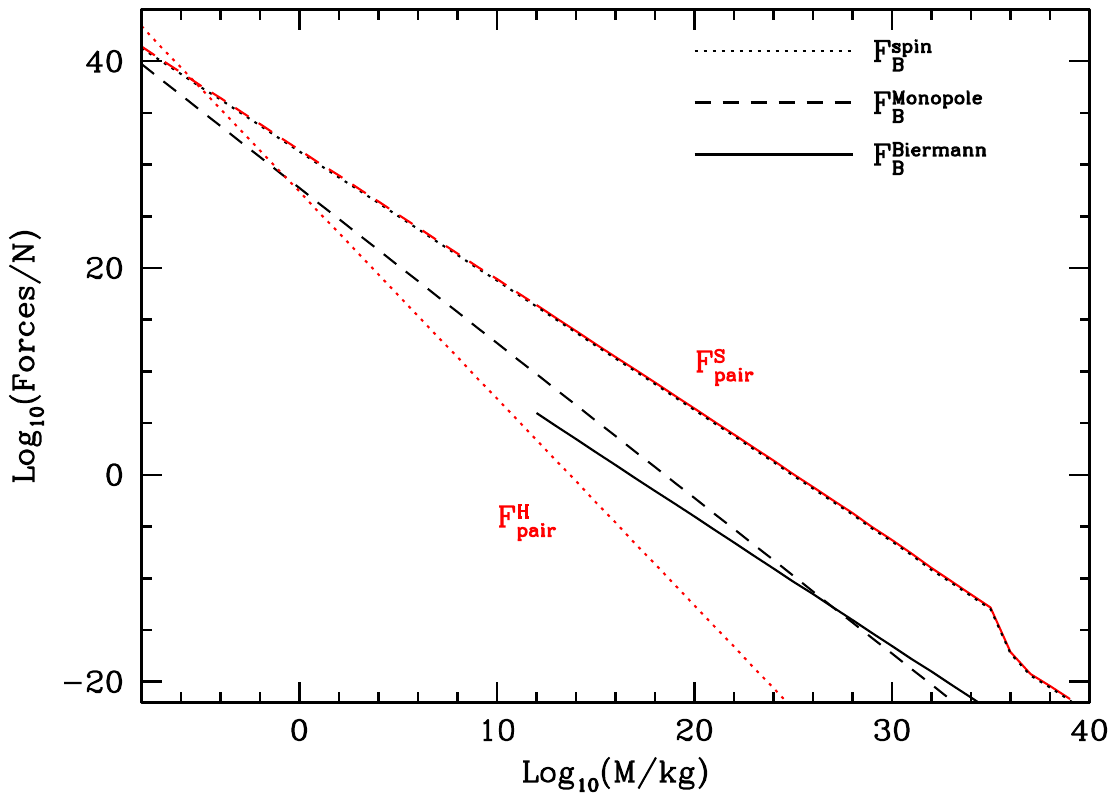}
    \vskip -8cm
    \caption{Forces associated to the processes of Hawking and Schwinger PBH discharge (red dotted and solid lines, respectively), and to the magnetic field produced by the Biermann battery process of high mass PBHs (black solid line), by charged maximally spinning PBHs (black dotted) and by Monopole charge (black dashed).} \label{fig:dischargeshield}
\end{figure}

In this case, instead of a deviation of trajectories, we will require the forces from  the PBH magnetic field to equal a "force equivalent" associated to the Hawking and Schwinger emission of charges which we approximate by $F\sim E/r_+$.  Namely,
\begin{equation}
    F_{\rm pair}^{H}\sim 3k_B T_{\rm H}/r_+,\label{eq:forceH}
\end{equation}
for the Hawking emission, and
\begin{equation}
    F_{\rm pair}^{S}\sim k_c e Q/r_+^2.\label{eq:forceS}
\end{equation}
for the Schwinger process.

These forces can then be compared to the maximal magnetic force for spinning PBHs assuming $v=c$ for the emitted pair considering the dipolar magnetic field of charged,  spinning black holes,
\begin{equation}
F_B^{\rm spin}(r_+)=\frac{a k_c Q e}{r_+^2}.
\end{equation}
It is immediately evident that this matches exactly the force equivalent for the Schwinger effect for $a=1$, providing an effective shielding of maximally spinning black holes against this process regardless of the PBH charge and mass.  

For PBHs that can form an accretion disc (see \cite{araya20}), namely with $M\gtrsim10^{12}$kg, the Biermann battery process can also produce a dipolar magnetic field with a force \cite{Biermann:1950}
\begin{equation}
F_B^{\text{Biermann}}(r_+)\simeq 2.3\times 10^{-3}  ec \left( \frac{M}{5M_\odot}\right)^{-9/4}\left( \frac{GM}{r_+^3}\right)^{-1/2} {\rm Tesla}/s.
\end{equation}

Lastly, the magnetic force associated to monopoles enclosed within the PBH horizon at formation reads \cite{araya20},
\begin{equation}
F_B^{\rm monopole}(r_+)=\frac{hc}{4\pi r_+^2}\left( \frac{\Omega_{{\rm mon},0}}{\Omega_{{\rm PBH},0}}\frac{M}{M_{\rm mon}}\right)^{1/2},
\end{equation}
where we adopt an energy density for monopoles within current upper limits $\Omega_{{\rm mon},0}=10^{-5}$ for a monopole mass $M_{\rm mon}=10^{16}\,\mbox{GeV}$, a candidate energy scale for the GUT transition, and the density parameter of PBHs reads $\Omega_{{\rm PBH},0}=f \Omega_{dm,0}$.

The comparison between these force equivalents can be seen in Fig. \ref{fig:dischargeshield} in red lines for the discharge processes and in black for the magnetic field forces.  Note that the Schwinger force equivalent is only valid for $M\gtrsim10^{12}$kg (solid red line) but we extend it to lower masses as a dashed line for illustrative purposes.

The magnetic force from maximally spinning PBHs (black dotted) is higher than both, the force associated to the Biermann battery (black solid), and the force due to magnetic monopoles in PBHs (black dashed).
$F_B^{\rm spin}$ is not only effective at shielding discharge by pair production via the Schwinger mechanism (red solid/dashed) but also by Hawking radiation (red dotted), for $M\gtrsim10^{-5}$kg PBHs.  Notice that this minimum mass is close to the lowest mass of PBHs that can form after our adopted reheating temperature, implying that the initial Poisson charge could be shielded by the magnetic field of Kerr extremal PBHs.  If we assume this argument to be valid for Planck mass PBHs, maximal Kerr-Newman black holes could also be stable under the Schwinger mechanism.

\section{Overview of the main results}
\label{sec:discussion}

In this work we argued whether black holes formed in the early Universe could hold electric charge at late times, independently of the fraction of dark matter in PBHs. For doing so, we studied the initial charge of PBHs as they enter the horizon, mechanisms of shielding against discharge due to magnetic fields sourced by the charged PBHs, and late time accretion of charges within dark matter haloes.

We provided estimates for the possible initial charge assuming Poissonian statistics for the abundance of charged particles in the Lagrangian volume of the PBH at the moment of their collapse.  When compared to the maximal Reissner-Nordstr\"om (RN) charge, the initial Poisson charge is between $2-10$ orders of magnitude smaller, with the larger differences present for the largest PBH masses explored here of $10^{40}$kg.  For PBHs to be candidates to DM, they should form as their volume enters into causal contact no earlier than the end of Inflation, for which we adopt a temperature of $10^{17}$GeV \cite{Carneiro:2021}.  This results in a lower mass for PBHs of $M\simeq10^{-4}$kg  with non-zero initial charge. 

\begin{figure}
    \centering
    \includegraphics[width=1\textwidth]{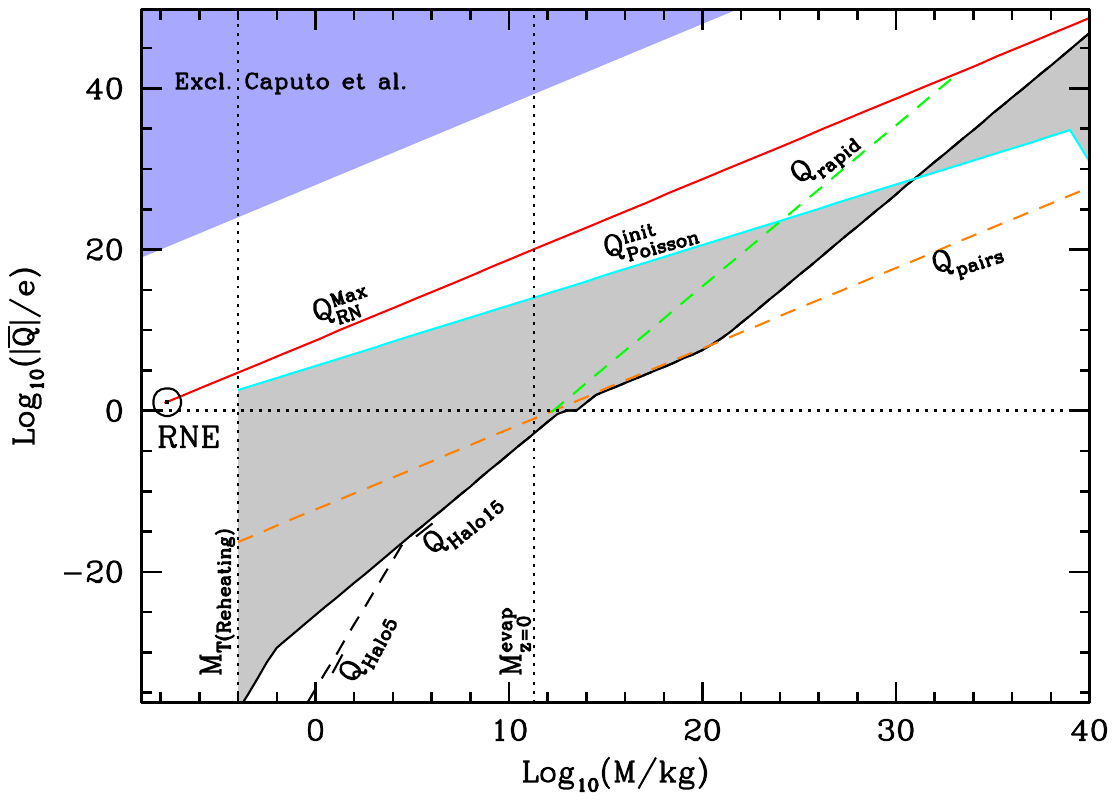}
    \vskip -8cm
    \caption{PBH charge in units of that of the electron corresponding to the Reissner Nordström maximal charge (red), the initial Poisson charge (cyan), and the present-day dark matter halo equilibrium charges for haloes of $M_h=10^{15}$ and $10^{5}M_\odot$ (black solid and dashed lines, respectively). The circle encloses the Planck mass extremal RN PBHs.  Dashed green and orange lines show the rapid and pair production charges for the Schwinger effect.  
    The horizontal dotted line shows the charge of one electron. The vertical dotted lines indicate the mass of PBHs that can form after reheating, and that evaporate via Hawking radiation by $z=0$, as indicated.  The shaded regions bracket possible values of the PBH charge taking into account emission mechanisms and late time values, with upper limits shielded by magnetic fields against accretion and discharge. The blue shaded area of the top left shows the limits excluded by \cite{Caputo:2019} for charged, although neutral on average, dark matter. } 
    \label{fig:final}
\end{figure}

We also explored the possibility of extreme RN PBHs formed during the reheating era after Inflation following \cite{Carneiro:2020,Carneiro:2021,Pigozzo:2021}, and showed that the required conditions could be met to make all of DM with such PBHs.  The volume of space taken up by the PBH formation is the same for these extremal Planck mass black holes as in the standard horizon crossing formation scenario for PBHs of this mass \cite{Sureda:2020}; this shows that these extremal PBHs could also contain a magnetic monopole charge. Furthermore, under the right conditions, the collision of the two electrons (or positrons) forming the PBH could leave a spin close to the extremal value for a Kerr BH, as the state resulting from two  $\frac{1}{2}-$spin particles can have a net spin of one (in units of $h/2\pi$), and the corresponding mass would be of about one Planck mass. These two conditions would  make the formation of an extremal BH more probable, as both magnetic and electric charges contribute to saturate the ``extremality'' condition of the resulting PBH, which would be of \textit{dyonic} Kerr-Newman type \cite{Semiz90}.  These conditions would also make the Planck mass extremal black hole stable under the Schwinger discharge (cf. Section \ref{sec:shielding}). 

Even though we show that accretion and emission processes could discharge these initial charges, we also show the possibility that these are shielded from discharge by magnetic fields.  The accretion of charges would be possible as PBHs are surrounded by a hot plasma in the early Universe, that could provide charges that nullify the PBH charge.  Also, PBHs could emit their excess charges via quantum production, either via Hawking Radiation or the Schwinger effect.  At late times, however, even if there are no shielding effects that allow PBHs to hold some of their initial charge, PBHs in haloes should be able to acquire charge.

We estimate the late times rate of accretion of electrons and protons within dark matter haloes which are long lived, and that for large enough masses contain an electron-proton plasma (alongside other chemical elements).  For low mass PBHs, $M\lesssim10^{13}$kg, the  rate is always lower than the Hubble rate at $z=0$ which allows us to ignore the continuity equation in the accretion process. When this rate is multiplied by the Hubble time it provides us with the fraction of PBHs with a charge of $Q=e$, or equivalently, with the charge per mass of PBHs.  The charge of PBHs above this mass would be limited by the amount of charge that can be accreted in a halo, during a Hubble time (taking into account the velocity of charged particles in the plasma), and can be as high as the charge that brings to a halt, at the outer horizon of the PBH, electrons incoming from a distance of one Debye length.  With these estimates we find that PBH charge increases steadily with mass.  For $M>10^{22}$ kg the final charge for PBHs of monochromatic mass $M$ in haloes reads, $Q/M\simeq1.15 \times 10^{-52}M\,C/\mbox{kg}$.

Therefore, as summarised in Fig. \ref{fig:final} our final charges  would lie between the negative { stable} halo charge and the negative initial Poisson charge (whichever is the largest, shown as gray shaded regions);  if the initial Poisson is positive, it can either be shielded and stay constant, or it can be totally discharged and later on, within dark matter haloes, it would tend toward the negative halo charge given by the black, solid or dashed lines, depending on the halo mass.
For comparison the figure also shows the  extremal RN charge. 

The later results do not contradict the well founded arguments pointing out that PBHs would quickly loose any initial electric charge. We simply find that this loss is not relevant { for PBHs living in present day haloes}, or that it can be shielded, and that a  robust charge should remain for PBHs of all masses, or for extremal Kerr-Newman PBHs of Planck mass. This implies that if PBHs constitute the DM for the Universe, they would most likely hold electric charge.

\section{Final remarks}
\label{sec:conclusions}
We have shown that if a non-zero fraction of the DM is made of PBHs, these would likely be charged, either by acquiring charge at their formation and maintaining it via shielding mechanisms, or by accreting charge from the plasma of present-day virialised haloes.

The effects of charged dark matter on different observables have been studied in the past by several authors, in particular given the possibility of DM particles with a small electric charge in extensions of the Standard Model of Particle Physics \cite{Munoz:2018}. Recent upper limits on the charge were given by \cite{Caputo:2019}, and if one were to naively extend these to the range of PBH masses here presented, even \textit{maximal} RN charges would be allowed, as the charge-per-mass in this case is rather small. This limit is in fact shown in Fig. \ref{fig:final}. However, the specific phenomenology associated to charged PBHs needs to be studied in order to place constraints on their charge and the possibility that they constitute the dark matter. For instance, charged PBHs would not only dynamically affect young star clusters, but could also electrically disrupt, { for instance,} molecular clouds. This is part of a forthcoming paper \cite{RubioXX}.

By comparing with constraints present in the literature regarding PBHs as $100\%$ of the dark matter, for monochromatic mass distributions current windows lie in the $M_{\text{\tiny{PBH}}}\sim 10^{15}\,\text{kg}$ and $M_{\text{\tiny{PBH}}}\sim M_{\odot}$ masses \cite{Carr:2020a,Sureda:2020}.  Our results show that for these monochromatic windows, the PBH charge would be of up to about $Q\sim 10^{-20}C/$kg, several orders of magnitude below current limits extrapolated from milli-charged DM studies \cite{Caputo:2019}.
For extended mass distributions, the average charge per mass should be obtained by weighting by the PBH mass function.

Finally, one very interesting possibility is the population of extremal (Kerr-Newman or dyonic Kerr-Newman) PBHs formed via collisions in the very early Universe, which would be characterised by $T_{\rm H}=0$ and would be shielded from the Schwinger effect by their magnetic field.  They would therefore escape evaporation constraints, and could conform the totality of the DM holding, in principle, the maximal spin and charge.

\acknowledgments

We thank Saulo Carneiro, Marco San Martín, Martín Makler and Nicholas Orlofsky for helpful discussions.  This project has received funding from the European Union's Horizon 2020 Research and Innovation Programme under the Marie Sk\l{}odowska-Curie grant agreement No 734374. MER acknowledges financial support provided under the European Union’s H2020 ERC Consolidator Grant “Gravity from Astrophysical to Microscopic Scales”, grant agreement no. GRAMS-815673 and INFN (Istituto Nazionale di Fisica Nucleare). IJA acknowledges funding from ANID, REC Convocatoria Nacional Subvenci\'on a Instalaci\'on en la Academia Convocatoria A\~no 2020, Folio PAI77200097. NDP acknowledges support from a RAICES grant from the Ministerio de Ciencia, Tecnología e Innovación, Argentina.

\appendix

\section{Shielding against accretion via magnetic monopole charge}
\label{app1}
If the PBH holds magnetic charge, as explored in \citep{araya20,BaiMag2020}, they would also produce additional radial magnetic field lines that could come into play after the dipolar field deviates, even if ever so slightly, incoming discharging particles from their originally radial trajectories, effectively shielding the PBH to discharge from its initial Poisson charge, $Q_0$ (Eq. (\ref{eq:Q0poisson})).

\begin{figure}
    \centering
    \includegraphics[width=0.47\textwidth]{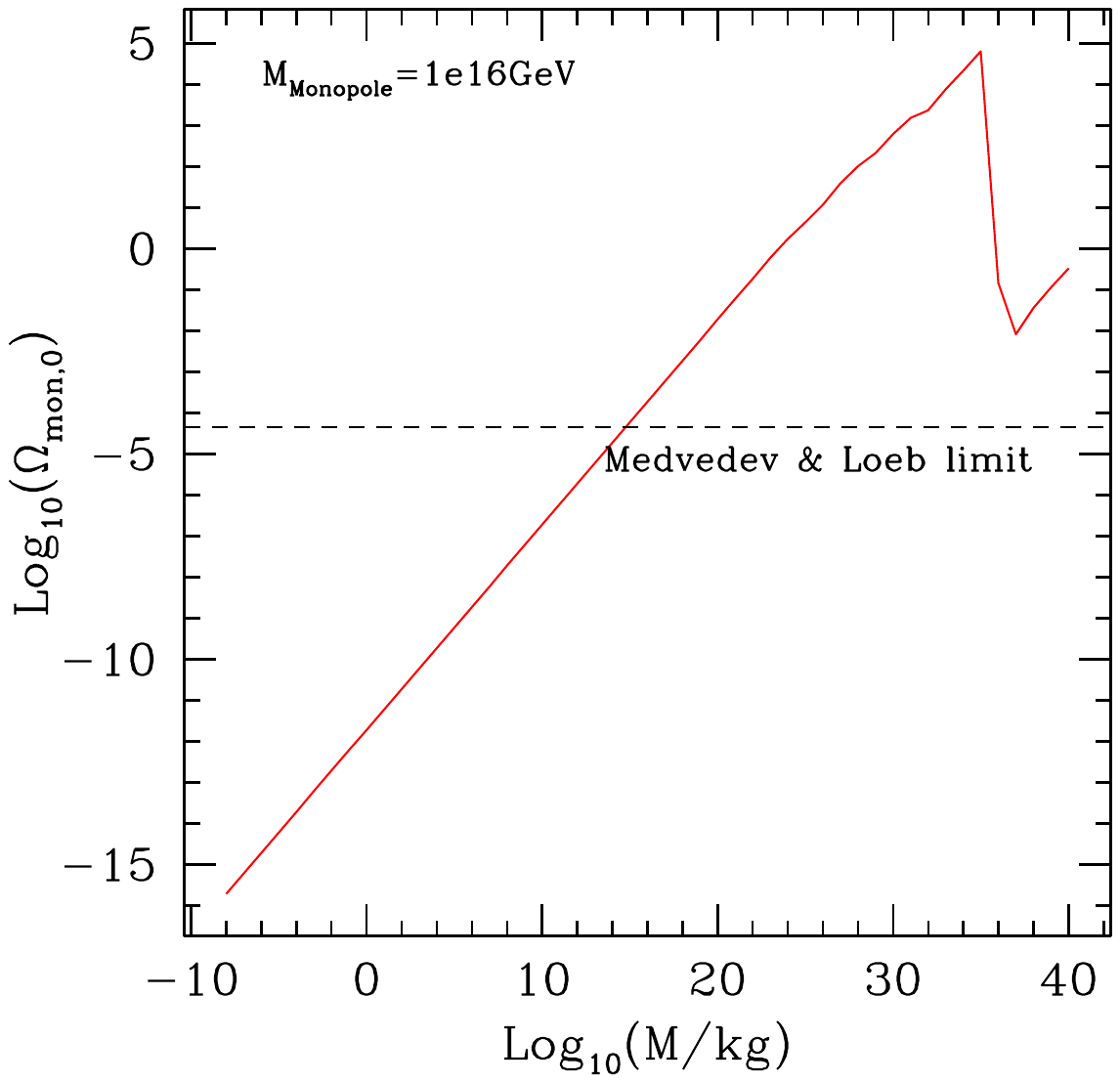}   \includegraphics[width=0.47\textwidth]{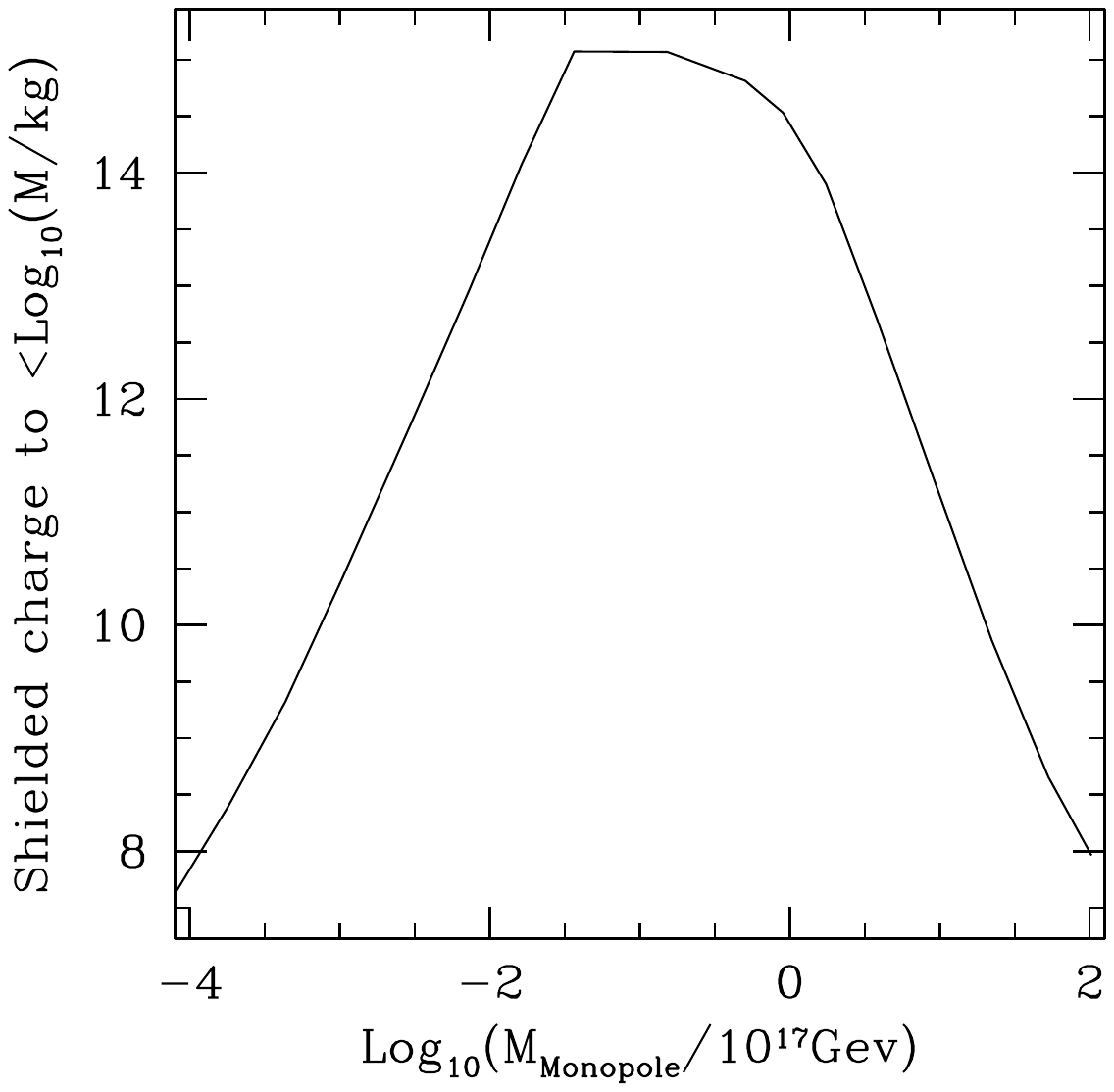}
    \vskip -2cm
    \caption{Left: Magnetic monopole density parameter needed for shielding PBHs against electrical discharge since their formation.  The horizontal line shows the upper limit for $M_{\rm GUT}=10^{16}\,\mbox{GeV}$. Right: Maximum PBH mass shielded from discharge of $Q_0$ for the case of rapidly spinning PBH $a\sim 1$ and with Poisson initial magnetic monopole charge.  This is shown as a function of the magnetic monopole mass.}
    \label{fig:ommon}
\end{figure}

For this to take place, the Lorentz force from the monopole magnetic field needs to be comparable to the electric one coming from the initial PBH charge. As shown in \cite{araya20}, the monopole charge is  proportional to the ratio of abundances of monopoles to PBHs, $\Omega_{\text{\tiny{mon}},0}/\Omega_{\text{\tiny{PBH}},0}$, which is assumed to stay constant throughout cosmic history. Both forces are comparable, assuming tangential motion, for
\begin{equation}
    \frac{\Omega_{\text{\tiny{mon}},0}}{\Omega_{\text{\tiny{PBH}},0}}=\left( \frac{4 \pi k_c Q_0 e }{hc} \right)^2 \frac{M_{\text{\tiny{GUT}}}}{M},\label{eq:ommon}
\end{equation}
where the magnetic monopole mass is taken to be the GUT mass scale  $M_{\text{\tiny{GUT}}}=10^{16}\,\text{GeV}$ \cite{Medvedev:2017JCAP}. The left panel of Fig. \ref{fig:ommon} shows the dependence of $\Omega_{\text{\tiny{mon}},0}$ with PBH mass, for the initial, Poissonnian PBH charge. Notice that the drop in $\Omega_{\text{\tiny{mon}},0}$ for $M \sim 10^{32}\,\text{kg}$ corresponds to the break in $Q_0$ due to the change of available charged particles in the Universe near the electron-positron annihilation energy scale (cf. Fig. \ref{fig:RN+Poiss}).  Given the upper limits for $\Omega_{\text{\tiny{mon}},0}$ for this assumed monopole mass, and according to \citep{Medvedev:2017JCAP}, PBHs of $M<10^{15}\,\text{kg}$ could in principle be shielded, being able to retain their initial Poisson charge.

The constraints  on $\Omega_{\text{\tiny{mon}},0}$ depend on the monopole mass. Using the corresponding bounds found in \citep{Medvedev:2017JCAP}, we can find the dependence of the mass of PBHs that could be shielded from discharge by the magnetic field produced by the monopoles contained within them. This is indeed shown in the right panel of Fig. \ref{fig:ommon}, and as can be seen, the population of PBHs with masses of up to $M=10^8 - 10^{15}\,\text{kg}$ could in principle be shielded from discharge, and thus retain their initial charge.

\bibliographystyle{unsrt}
\bibliography{references}
\end{document}